\pdfoutput=1
\documentclass[11pt,twoside,a4paper,cmspaper,final,collab]{cms-tdr}
\def\svnVersion{2bacb58}\def\svnDate{2025/08/19}\def\cmsCernNoTag{CERN-EP-2024-174}\def\cmsCernDate{\today}\def\cmsMessage{Published in Physical Review D as \href{http://dx.doi.org/10.1103/95q6-vvlp}{\doi{10.1103/95q6-vvlp}.}}
\begin{document}\cmsNoteHeader{TOP-22-002}

\newcommand{\fullLumi}{138\fbinv}
\newcommand{\wjets}{\ensuremath{\PW+\text{jets}}\xspace}
\newcommand{\zjets}{\ensuremath{\PZ+\text{jets}}\xspace}
\newcommand{\hut}{\ensuremath{\PH\PQu\PQt}\xspace}
\newcommand{\hct}{\ensuremath{\PH\PQc\PQt}\xspace}
\newcommand{\hqt}{\ensuremath{\PH\PQq\PQt}\xspace}
\newcommand{\pp}{\ensuremath{\Pp\Pp}\xspace}
\newcommand{\tuH}{\ensuremath{\PQt \to \PH\PQu}\xspace}
\newcommand{\tcH}{\ensuremath{\PQt \to \PH\PQc}\xspace}
\newcommand{\brHut}{\ensuremath{\mathcal{B}(\PQt \to \PH\PQu)}\xspace}
\newcommand{\brHct}{\ensuremath{\mathcal{B}(\PQt \to \PH\PQc)}\xspace}
\newcommand{\gjet}{\ensuremath{\PGg+\text{jets}}\xspace}
\newcommand{\ww}{\ensuremath{\PW\PW}\xspace}
\newcommand{\zz}{\ensuremath{\PZ\PZ}\xspace}
\newcommand{\wz}{\ensuremath{\PW\PZ}\xspace}
\newcommand{\fr}{\ensuremath{\epsilon_{\mathrm{TL}}}\xspace}
\newcommand{\flipr}{\ensuremath{\epsilon_{\mathrm{q}}}\xspace}
\newcommand{\mur}{\ensuremath{\mu_{\mathrm{R}}}\xspace}
\newcommand{\muf}{\ensuremath{\mu_{\mathrm{F}}}\xspace}
\newcommand{\khut}{\ensuremath{\kappa_{\hut}}\xspace}
\newcommand{\khct}{\ensuremath{\kappa_{\hct}}\xspace}
\newlength\cmsTabSkip\setlength{\cmsTabSkip}{1ex}
\ifthenelse{\boolean{cms@external}}{\providecommand{\cmsTable}[1]{#1}}{\providecommand{\cmsTable}[1]{\resizebox{\textwidth}{!}{#1}}}
\ifthenelse{\boolean{cms@external}}{\providecommand{\cmsLeft}{upper\xspace}}{\providecommand{\cmsLeft}{left\xspace}}
\ifthenelse{\boolean{cms@external}}{\providecommand{\cmsRight}{lower\xspace}}{\providecommand{\cmsRight}{right\xspace}}

\cmsNoteHeader{TOP-22-002}
\title{Search for flavor-changing neutral current interactions of the top quark mediated by a Higgs boson in proton-proton collisions at \texorpdfstring{13\TeV}{13 TeV}}

\date{\today}

\abstract{A search for flavor-changing neutral current interactions of the top quark (\PQt) and the Higgs boson (\PH) is presented. The search is based on proton-proton collision data collected in 2016--2018 at a center-of-mass energy of 13\TeV with the CMS detector at the LHC, and corresponding to an integrated luminosity of 138\fbinv. Events containing a pair of leptons with the same-sign electric charge and at least one jet are considered. The results are used to constrain the branching fraction ($\mathcal{B}$) of the top quark decaying to a Higgs boson and an up (\PQu) or charm (\PQc) quark. No significant excess above the estimated background was found. The observed (expected) upper limits at 95\% confidence level are found to be 0.072\% (0.059\%) for $\mathcal{B}(\PQt\to \PH\PQu)$ and 0.043\% (0.062\%) for $\mathcal{B}(\PQt\to \PH\PQc)$. These results are combined with two other searches performed by the CMS Collaboration for flavor-changing neutral current interactions of top quarks and Higgs bosons in final states where the Higgs boson decays to either a pair of photons or a pair of bottom quarks. The resulting observed (expected) upper limits at 95\% confidence level are 0.019\% (0.027\%) for $\mathcal{B}(\PQt\to \PH\PQu)$ and 0.037\% (0.035\%) for $\mathcal{B}(\PQt\to \PH\PQc)$.}

\hypersetup{
pdfauthor={CMS Collaboration},
pdftitle={Search for flavor-changing neutral current interactions of the top quark mediated by a Higgs boson in proton-proton collisions at sqrt(s) = 13 TeV},
pdfsubject={CMS},
pdfkeywords={CMS, FCNC}}

\maketitle 

\section{Introduction}
\label{sec:intro}

Quark decays mediated by flavor-changing neutral currents (FCNCs) are forbidden at tree level in the standard model (SM) and are suppressed at higher orders in the perturbative expansion by the Glashow--Iliopoulos--Maiani mechanism~\cite{Glashow:1970gm} and Cabibbo--Kobayashi--Maskawa unitarity constraints~\cite{Kobayashi:1973fv}.
As a result, the SM branching fractions for the decay of a top quark (\PQt) into a Higgs boson (\PH) and an up quark (\PQu), \tuH, or a charm quark (\PQc), \tcH, are expected to be of the order of $10^{-17}$ and $10^{-15}$, respectively~\cite{Eilam:1990zc,Mele_1998,AguilarSaavedra:2004wm,Zhang_2013}.

Many scenarios of physics beyond the SM may enhance these interactions, possibly by many orders of magnitude.  
Examples of such models include those of warped extra dimensions~\cite{Azatov_2009},
composite Higgs boson models~\cite{Azatov_2014},
two-Higgs-doublet models (2HDMs)~\cite{Bejar:2000ub,Guasch_1999,Cao_2007,Cao_2014}
including supersymmetric models with {\it R}-parity violation~\cite{Eilam_2001}, 
and quark-singlet models~\cite{AguilarSaavedra:2002kr}.
These scenarios may lead to sizable FCNC interactions of quarks with any of the neutral force mediators or with the Higgs boson.
The \hct interaction in particular may be enhanced in 2HDMs~\cite{Hou:1991un,Baum:2008qm,Kao:2011aa,Altunkaynak:2015twa}, including scenarios of flavor-violating Yukawa couplings~\cite{Chen:2013qta}.  

Searches for FCNC interactions of the top quark mediated by a Higgs boson have been performed by the ATLAS~\cite{Aaboud:2017mfd,Aaboud:2018pob,Aaboud:2018oqm,ATLAS:2024mih} and CMS~\cite{CMS:2016obj,Sirunyan:2017uae,CMS:2021gfa,CMS:2021cqc} Collaborations.
This paper presents searches for FCNC top quark interactions in a final state having at least two leptons ($\Pe,\PGm$) with the same-sign (SS) electric charge and at least one jet. 
The CMS Collaboration previously reported a search in a similar final state, analyzing 19.7\fbinv of 8\TeV proton-proton ($\Pp\Pp$) collision data~\cite{CMS:2016obj}. 
The results presented here are obtained from the analysis of ($\Pp\Pp$) collisions at a center-of-mass energy of 13\TeV, targeting the decay of the Higgs boson to \ww, \zz, or $\PGt\PGt$ leading to the final states described above.
The data were collected with the CMS detector at the CERN LHC in 2016--2018, and correspond to an integrated luminosity of \fullLumi.  

\begin{figure} [htb!]
    \centering
    \includegraphics[width=0.35\textwidth]{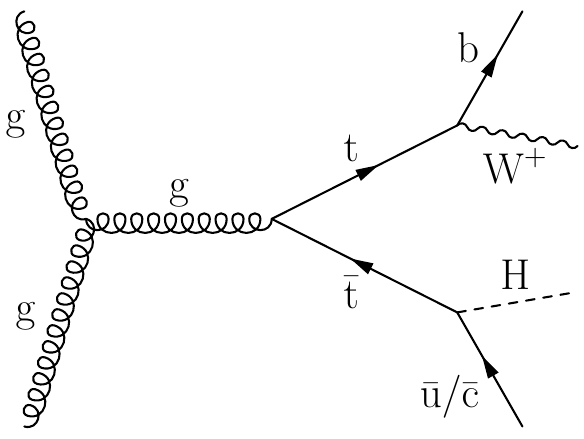}
    \includegraphics[width=0.35\textwidth]{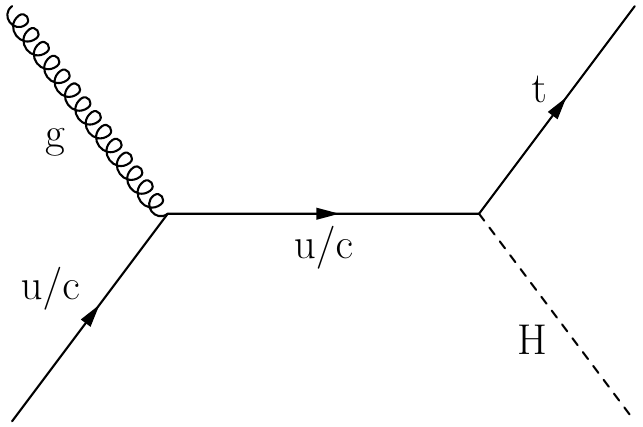}
    \caption{Representative Feynman diagrams for the production modes considered:  \ttbar production with the FCNC decay of the top quark to a Higgs boson and an up or charm quark (TT, \cmsLeft), and FCNC-associated production of a single top quark with a Higgs boson (ST, \cmsRight).
    }
    \label{fig:feyn}
\end{figure}

Two channels for signal production are considered, the decay of a top quark to a Higgs boson and an up or charm quark in top quark-antiquark (\ttbar) pair production (TT production mode), as well as the associated production of a single top quark (ST production mode) with a Higgs boson via an up or charm quark, as shown in Fig.~\ref{fig:feyn}. 
The anomalous \hqt couplings enter through the top quark decay in the TT mode, while appearing in the matrix elements of the ST production mode with the single top quark subsequently decaying as normal, i.e. to a bottom quark and \PW boson.
Other production modes are possible but are sub-dominant, such as the $\PQt\PW$ production mode which accounts for less than 4\% of the signal.  
For the sake of consistency with other published searches for this phenomenon, only the TT and ST modes shown in Fig.~\ref{fig:feyn} are considered.
Results from a search in final states with leptons are combined with other FCNC searches performed by the CMS Collaboration targeting the decay of the Higgs boson to a pair of b quarks~\cite{CMS:2021gfa} or a pair of photons~\cite{CMS:2021cqc}.

\section{The CMS detector}
\label{sec:CMS}

The central feature of the CMS apparatus is a superconducting solenoid of 6\unit{m} internal diameter, providing a magnetic field of 3.8\unit{T}. Within the solenoid volume are a silicon pixel and strip tracker, a lead tungstate crystal electromagnetic calorimeter (ECAL), and a brass and scintillator hadron calorimeter, each composed of a barrel and two endcap sections. Forward calorimeters extend the pseudorapidity ($\eta$) coverage provided by the barrel and endcap detectors. Muons are measured in gas-ionization detectors embedded in the steel flux-return yoke outside the solenoid. A more detailed description of the CMS detector, together with a definition of the coordinate system used and the relevant kinematic variables, can be found in Ref.~\cite{CMS:2008xjf}.

Events of interest are selected using a two-tiered trigger system. The first level trigger, composed of custom hardware processors, uses information from the calorimeters and muon detectors to select events at a rate of around 100\unit{kHz} within a fixed latency of about 4\mus~\cite{CMS:2020cmk}. The second level trigger, known as the high-level trigger, consists of a farm of processors running a version of the full event reconstruction software optimized for fast processing, and reduces the event rate to around 1\unit{kHz} before data storage~\cite{CMS:2016ngn}.

\section{Simulated samples}
\label{sec:mc}
Samples of simulated events are used in the design and validation of the analysis, the training and optimization of event classifiers, and in the estimate of expected yields from signal and rare SM background processes.
Three separate sets of simulated events for each process are used in order to match the different data-taking conditions and algorithms used in 2016, 2017, and 2018.

The \POWHEG{}~v2 event generator~\cite{Nason:2004rx,Frixione:2007vw,Alioli:2010xd} is used to simulate \ttbar production~\cite{Frixione:2007nw}, \wz and \zz pair production~\cite{Nason:2013ydw}, \PH production via gluon fusion~\cite{Bagnaschi:2011tu}, and \ttbar{}\PH{} production~\cite{Hartanto:2015uka} at next-to-leading order (NLO) in perturbative quantum chromodynamics (QCD).
The \textsc{JHUgen} generator~\cite{Bolognesi:2012mm} is used to simulate the decay of the Higgs boson into vector bosons in the gluon fusion production mode.
Events with single top quarks produced in association with W bosons ($\PQt\PW$) are simulated at NLO with \POWHEG{}~v1~\cite{Re:2010bp} (2016) or \POWHEG{}~v2 (2017--2018).
Samples of the Drell--Yan and \PW boson production processes are generated with up to four additional partons in the matrix element (ME) calculations using the \MGvATNLO event generator at leading order (LO)~\cite{Alwall:2014hca}.
The \MGvATNLO{}~v2.3.3 (v2.4.2) package is used for samples corresponding to 2016 (2017--2018) data-taking conditions.
The production of $\PW^{\pm}\PW^{\pm}$, as well as the $\PQt\PW\PZ$, $\ttbar\PGg$, and $\ttbar+\PX\PX$ (with $\PX = \PW, \PZ, \PH$) processes, are generated using \MGvATNLO at LO.
Triple vector boson production ($\PV\PV\PV$, $\PV = \PZ,\PW$), diboson production in association with a photon ($\PW\PZ\PGg$, $\PW\PW\PGg$) as well as the $\ttbar\PZ$, $\PQt\PZ\Pq$, $\PQt\PGg$ and $\ttbar\ttbar$ processes are simulated using \MGvATNLO{} at NLO in QCD.
The $\ttbar\PW$, $\PW\PGg$, $\PZ\PGg$, and $\PV\PH$ processes are simulated at NLO with up to one or two extra partons in the ME calculations using \MGvATNLO.

{\tolerance=800
The NNPDF3.0~\cite{Ball:2014uwa} LO and NLO parton distribution function (PDF) sets are used to generate LO and NLO samples with 2016 data-taking conditions, while NNPDF3.1 next-to-NLO (NNLO)~\cite{Ball:2017nwa} is used for 2017--2018 samples.
Event generators are interfaced with
\PYTHIA~v8.226~\cite{Sjostrand:2014zea} 
using the CUETP8M1 tune~\cite{Skands:2014pea,Khachatryan:2015pea} for 2016 and \PYTHIA~v8.230 using the CP5 tune~\cite{Sirunyan:2019dfx} for 2017--2018 to simulate the parton shower, fragmentation, and hadronization of initial- and final-state partons, along with the underlying event.
The MLM~\cite{Alwall:2007fs} and \textsc{FxFx}~\cite{Frederix:2012ps} prescriptions are employed to remove double counting of additional partons generated with \MGvATNLO and \PYTHIA for the LO and NLO samples, respectively.
The \GEANTfour toolkit~\cite{Agostinelli:2002hh} is used to model the response of the CMS detector.
\par}

The effective Lagrangian with an FCNC coupling is:
    \begin{equation}
        \begin{aligned}
            \mathcal L = \sum_{\PQq=\PQu,\PQc} \frac{g}{\sqrt{2}} \PAQt\kappa_{\hqt} \Big( F_{\PH\PQq}^{\mathrm{L}} P_{\mathrm{L}} + F_{\PH\PQq}^{\mathrm{R}} P_{\mathrm{R}} \Big) \PQq\PH + \text{h.c.}. \label{eqn:lagrangian}
        \end{aligned}
    \end{equation}
Here, $P_L (P_R)$ are the left-hand (right-hand) projection operators defined as $P_L =(1-\gamma_5)/2$ and $P_R = (1+\gamma_5)/2$, and the notation h.c. indicates that the Hermitian conjugate terms are to be included in the sum. 
It is implemented in the \textsc{FeynRules} package~\cite{ALLOUL20142250}, and the universal \textsc{FeynRules} output format~\cite{Degrande_2012} is used to generate the signal model.
The complex chiral parameters are arbitrarily fixed to the values $F_{\PH\PQq}^{\mathrm{L}} = 1$ and $F_{\PH\PQq}^{\mathrm{R}} = 0$, as these parameters have no impact on the signal cross sections and usually have a small impact on kinematic distributions of objects in the event.
Signal samples are generated in the TT and ST production modes assuming, for each mode, one of the two $\kappa$ couplings is non-zero at a time. 
For the ST sample, the production and decays of the top quark and antiquark are simulated using \MGvATNLO{} at LO.
The ST production mode cross section is calculated at LO precision with \MGvATNLO~v2.6.0 as 72.6 (10.0)\unit{pb}, equivalent to a coupling constant \khut = 1 (\khct = 1).
The difference of the respective production cross sections originates from the up and charm quark parton distribution functions of the initial state protons.
The TT production signal sample is generated with \MGvATNLO{}~v2.4.2  at LO with up to two extra partons in the ME.
The TT production mode cross section is taken to be 832\unit{pb}, as calculated at NNLO precision in perturbative QCD including soft-gluon resummation to next-to-next-to-leading-logarithm~\cite{Czakon_2014}.
The SM and FCNC decays of the top quark and antiquark are simulated using \textsc{MadSpin}~\cite{Artoisenet:2012st}.
The top quark and Higgs boson masses are set to 172.5 and 125.0\GeV, respectively.
The Higgs boson decay is simulated using \PYTHIA~v8.226 (2016) or \PYTHIA~v8.230 (2017 and 2018).
The generated signal samples are filtered to have one leptonically decaying top quark and one Higgs boson decaying to $\PW\PW$, $\PZ\PZ$, or $\PGt\PGt$.
The cross sections applied in the analysis account for the branching fraction of the filtered LO decays.
The adjusted cross sections are 12.08\unit{pb} for the TT production and 7.32 (1.01)\unit{pb} for the ST \hut (\hct) production.

Additional simulated minimum bias $\Pp\Pp$ interactions within the same or adjacent bunch crossings (pileup) are included 
in the simulated events, and events are reweighted according to the observed instantaneous luminosity and the total inelastic pp cross section of 69.2\unit{mb}~\cite{CMS:2020ebo}.

\section{Event selection and search strategy}
\label{sec:evtsel}
This analysis considers events with at least two leptons (electron or muon) with the SS electric charge and at least one jet.
The main backgrounds are from detector effects (namely nonprompt leptons and charge misidentified leptons) and SM processes that produce an SS lepton pair, including processes producing multiple gauge bosons and/or top quarks. 
Backgrounds are suppressed by making requirements on the kinematic variables of reconstructed physics objects, the same as those presented in Refs.~\cite{SS:2016paper,SS:2020paper,fourtoppaper}.

Events are reconstructed using the particle-flow (PF) algorithm~\cite{Sirunyan:2017ulk}, which combines information from the CMS subdetectors to identify charged and neutral hadrons, photons, electrons, and muons, collectively referred to as PF candidates.
These candidates are associated with reconstructed vertices.
The primary vertex is taken to be the vertex corresponding to the hardest scattering in the event, evaluated using tracking information alone, as described in Section 9.4.1 of Ref.~\cite{CMS-TDR-15-02}.
The physics objects used include jets clustered from PF candidates associated with the PV and the magnitude of missing transverse momentum, \ptvecmiss~\cite{Sirunyan:2019kia}.
The \ptvecmiss is computed as the negative vector sum of the transverse momentum (\ptvec) of all the PF candidates, excluding charged-hadron candidates that do not originate from the PV.
The magnitude of this vector is referred to as \ptmiss.

Electron candidates are reconstructed by combining clusters of energy deposits in the ECAL with tracks~\cite{Sirunyan:2020ycc}.
The electron identification is performed using shower shape variables, track-cluster matching variables, and track quality variables.
To reject electrons originating from photon conversions inside the detector, electron candidates are required to have at most one missing measurement in the innermost tracker layers and to be incompatible with any conversion-like secondary vertices.
Muon candidates are reconstructed by geometrically matching tracks from measurements in the muon system and tracker, and fitting them to form a global muon track.
Good muon candidates are selected using the qualities of the geometrical matching and of the track~\cite{Sirunyan:2018fpa}.

Selected electrons (muons) are required to have $\pt>20\GeV$, $\abs{\eta}<2.4\: (2.5)$ and originate from the PV.
Additionally, electrons in the transition region $1.442 < \abs{\eta} < 1.556$, between the barrel and endcaps of the ECAL, are excluded since the reconstruction of an electron object in this region is not optimal.
Leptons are required to be isolated using a logical combination of three isolation variables: the energy surrounding the lepton candidate in a fixed cone size $\Delta R=0.4$, the energy surrounding the lepton candidate in a cone size that depends on the candidate \pt, and the transverse momentum of the lepton candidate relative to the residual momentum of the nearest jet after lepton momentum subtraction. 
This definition is designed to distinguish leptons produced in decays of \PW or \PZ bosons (``prompt leptons") from leptons produced in hadron decays, in conversions of photons in jets, or hadrons misidentified as leptons (``nonprompt leptons").
Lepton selection efficiencies are in the ranges of 45--80 (70--90)\% for electrons (muons), increasing as a function of the lepton \pt and reaching a maximum value for $\pt > 60\GeV$. 
Details of the lepton isolation requirements can be found in~\cite{SS:2016paper}.

Hadronic jets are clustered from neutral and charged PF candidates associated with the PV, using the anti-\kt algorithm~\cite{Cacciari:2008gp,Cacciari:2011ma} and a distance parameter of 0.4.
The jet momentum is determined as the vector sum of all PF candidate momenta in the jet.
Jet energy corrections are derived from simulation studies so that the average measured energy of jets becomes identical to that of particle level jets.
In situ measurements of the momentum balance in dijet, \gjet, \zjets, and multijet events are used to determine any residual differences between the observed and simulated jet energy scale, and appropriate corrections are made~\cite{CMS:2016lmd}. 
Additional selection criteria are applied to each jet to remove jets potentially dominated by instrumental effects or reconstruction failures.
The \ptvecmiss is modified to account for corrections to the energy scale of the reconstructed jets in the event.
The \HT is defined as the scalar \pt sum of jets in an event.

Selected jets must have $\pt>25\GeV$, $\abs{\eta}<2.4$, and be separated from isolated leptons by $\Delta R = \sqrt{\smash[b]{(\Delta\eta)^2+(\Delta\phi)^2}}> 0.4$, where $\phi$ is the azimuthal angle in radians.
Jets containing the decay of a \PQb hadron are identified as bottom quark jets (\PQb-tagged jets) using a deep neural network algorithm, \textsc{DeepJet}~\cite{Sirunyan_2018,Bols:2020bkb,CMS-DP-2018-058}.
The discriminator working point is chosen so that the misidentification rate to tag light-flavor or gluon jets is approximately 1\%.
This choice results in an efficiency to identify a bottom quark jet in the range 75--85\% for jets with \pt of 25--400\GeV, and a misidentification rate of about 15\% for jets originating from a charm quark.
Jets failing the \PQb tagging requirement are further required to have $\pt > 30\GeV$ to improve signal purity.
The \textsc{DeepJet} algorithm is also used to identify jets containing the decay of a \PQc-flavored hadron (\PQc-tagged jets)~\cite{CMS:2021scf}.
The DeepJet algorithm is a multiclassifier, allowing for a \PQb tagging discriminator score and a \PQc tagging discriminator score to be returned by the same algorithm.
Jets arising from \PQb and \PQc hadrons tend to be amongst the hardest jets in the event and thus the \PQb and \PQc tagging discriminator scores for the leading three jets are used for event classification.

Events containing at least three leptons, referred to as multilepton events, or a pair of leptons with the same-sign electric charged are selected using a logical OR of triggers requiring the presence of two isolated leptons.
The trigger efficiency for electrons is 90--98\%, reaching its maximum value for $\pt>30\GeV$.
For muons, the trigger efficiency is around 92\%.
The uncertainty in the trigger efficiency measurement is 1--5\% per trigger lepton.
The trigger requirements are summarized in Table~\ref{tab:trigs}.
\begin{table}
\centering
	\topcaption{Summary of the trigger thresholds used to select the analysis data set.
	The triggers require two isolated leptons, with different \pt requirements depending on their flavors.
        For each trigger, the  \pt threshold 1 corresponds to the lepton listed first, and the \pt threshold 2 to the lepton listed last.}	
	\begin{scotch}{ ccc }
		Trigger			& $p_{\mathrm{T,1}}$ threshold	& $p_{\mathrm{T,2}}$ threshold \\
		\hline
        \Pe{}\Pe{}      & 23\GeV               & 12\GeV \\
		\Pe{}\PGm 		& 23\GeV				& 8\GeV \\
		\PGm{}\Pe 		& 23\GeV				& 12\GeV \\
        \PGm{}\PGm{}    & 17\GeV               & 8\GeV \\
    	\end{scotch}
	\label{tab:trigs}
\end{table}
\begin{table*}
\centering
	\topcaption{
	Baseline analysis selections.	
	}
	\begin{scotch}{ ll}
		Physics objects & Selection criteria\\
		\hline
		\multirow{2}{*}{Lepton} 				& Pair of isolated SS leptons (\Pe or \PGm),\\
												& lead $\pt > 25\GeV$, else $\pt > 20\GeV$, $\abs{\eta_{\Pe}}<2.4$, $\abs{\eta_{\PGm}}<2.5$ \\ [\cmsTabSkip]
		\multirow{2}{*}{Jet} 					& $\ge$2 in SS events or $\ge$1 in multilepton events, \\
												& $\pt > 30\GeV$, $\abs{\eta}<2.4$ \\ [\cmsTabSkip]
		\PQb-tagged jet							& $\pt > 25\GeV$, $\abs{\eta}<2.4$\\
		$m_{\Pell\Pell}$(SF)						& $>$12\GeV \\
		$m_{\Pell\Pell}$(any flavor, any charge)	& $>$8\GeV \\
		$m_{\Pe\Pe}$(SS)	&  $<$75 or $>$105\GeV \\
	\end{scotch}
	\label{tab:sels}
\end{table*}

Events are vetoed if any pair of leptons in the events is consistent with coming from a known resonance. 
Events with a pair of leptons with an invariant mass less than 8 GeV are vetoed, irrespective of lepton flavor or charge. 
Additionally, events with a pair of same-flavor leptons of any charge are vetoed if the invariant mass of the pair is greater than 12\GeV. 
Finally, events with a pair of electrons with the same-sign electric charge are vetoed if the invariant mass of the pair is between 75 and 105\GeV.  
A baseline analysis selection defined in terms of requirements on these objects is summarized in Table~\ref{tab:sels}. 
The main SM processes that contribute at this stage of the selection are dibosons, associated production of gauge bosons with top quarks, \ttbar, \wjets, and Drell--Yan.

Events passing the baseline selection are classified using a boosted decision tree (BDT) trained on simulated samples of signal and background processes. 
Training is performed using a mixture of the two signal prodution modes weighted according to the cross sections in Section~\ref{sec:mc}.
Two BDTs are trained: one for identifying the \hut signal and one the \hct signal.
Both BDTs are trained using the \textsc{xgboost}~\cite{xgboost_2016} framework with 33 input features, which were selected for their expected differences between signal and background events:
\begin{itemize}
	\item{The number of electrons, jets, and \PQb-tagged jets in the event,}	
	\item{kinematic features of up to three leptons including the \pt, $\abs{\eta}$, transverse and longitudinal impact parameters, and the transverse mass of the lepton and the \ptmiss,}
	\item{features of the leading three jets: \pt, \PQb and \PQc tagging discriminator score,}
	\item{invariant mass of the two \pt-leading leptons,}
	\item{\pt of the jet with the largest $\abs{\eta}$,}
	\item{\PQb tagging discriminator score of the \pt-leading \PQb-tagged jet,}
        \item{\pt of the jet with the highest \PQb tagging discriminator score,}
	\item{\ptmiss,}
	\item{\HT.}
\end{itemize}

The input features above are primarily numbers, kinematics, and other properties of selected leptons and jets. 
These are among the most important discriminating variables between signal and backgrounds that are nearly entirely from processes containing one or more top quarks.
Separate trainings for the ST and TT signal modes were evaluated and did not yield significant improvement.

Search regions (SRs) are defined by bins in the BDT discriminator value.
Bin edges are determined using the \textsc{scikitlearn}~\cite{scikit-learn} quantile transformation function so that the signal acceptance is approximately uniform across SRs, reducing the effects of outlier events.
Expected yields of the signal process in these SRs are determined from simulation after applying scale factors (SFs) that correct for differences between data and simulation in the efficiencies of the applied triggers, lepton selection, and \PQb tagging. 

\section{Background estimation}
\label{Sec:BkgEst}

Events containing an SS lepton pair and jets arise from several SM processes.  
The first category consists of events that contain one or more nonprompt leptons.
Events containing a pair of prompt leptons with the SS electric charge arise from SM processes including diboson and triboson production and associated production of a boson with a \ttbar.  
This background is subdominant and is estimated from simulation.  
The smallest category of background consists of dilepton events from \ttbar, Drell--Yan, and \ww processes with a charge-misidentified prompt electron.  
Descriptions of the methods used to estimate each category of background and the validation studies follow.

Semileptonic \ttbar and leptonic \wjets processes with an additional nonprompt lepton with the same-sign electric charge as a prompt lepton constitute the largest background in this search.
This background is estimated using the ``tight-to-loose'' (TL) ratio method~\cite{SS:2016paper}.
The tight identification and isolation requirements used to select prompt leptons are relaxed to define a loose lepton selection that is enriched in nonprompt leptons.
The efficiency, \fr, for a nonprompt lepton that satisfies the loose selection to also pass the tight requirement is measured in a control sample of single-lepton events, as a function of the lepton flavor, \pt, and $\abs{\eta}$.  
Contamination from processes producing a prompt lepton, mostly \wjets and followed by Drell--Yan, is subtracted using the simulation. 
The loose-lepton definition was optimized to achieve an \fr that is approximately the same for lepton candidates arising from heavy or light flavor hadrons.  
Furthermore, for leptons that fail the tight selection, the lepton \pt is redefined as the sum of the lepton \pt and the energy in the isolation cone, where the isolation cone is an angular distance $\Delta R =0.4$ from the lepton \pt.  
The redefinition accounts for the momentum of the parent parton. Together, the criteria allow constructing an \fr that is approximately independent of sample and SR.
See Ref.~\cite{SS:2016paper} for further details.
For each SR, a control region (CR) is constructed with the same selection criteria as the SR but requiring that at least one lepton passes the loose selection but fails the tight requirement.  
The number of events with a nonprompt lepton in each SR is estimated by weighting each loose-not-tight lepton in a corresponding CR event by a factor of $\fr / (1 - \fr)$.

The simulation is used to evaluate the performance of the method.  
A TL efficiency, measured from simulated samples of multijet production, is used to predict the number of events with a nonprompt lepton expected to enter the SRs.  
The estimate is compared with the observed number of events with a nonprompt lepton in simulated \ttbar and \wjets samples.  
The predicted and observed rates of events with a nonprompt lepton are compared in each SR and as functions of kinematic properties.
These rates are found to agree within 30\%, which is taken as a flat systematic uncertainty in the final nonprompt-lepton background estimate.

The second category of background consists of SM processes that produce a pair of prompt SS leptons.
Diboson and triboson processes and associated production of a boson with \ttbar are the most significant processes in this background category.
Smaller contributions arise from processes including a genuine photon such as $\PW\PGg,\PZ\PGg,\PQt\PGg$ via photon conversion, and ``rare'' processes, such as $\PQt\PW\PZ$ and \ttbar\ttbar.

As it is not possible to construct a CR enriched in these processes with a large number of events and high purity, the backgrounds are estimated from simulation. 
Expected yields are estimated after applying SFs that account for small differences between data and simulation in the measured trigger, lepton selection, and b tagging efficiencies, with associated systematic uncertainties described in Section~\ref{sec:uncertainties}.

The smallest category of background consists of dilepton \ttbar and Drell--Yan events where the charge of an electron is misidentified, thus misclassifying a pair of prompt leptons with opposite-sign electric charge as an SS pair.
Referred to as the ``charge flip'' background, it is estimated using a method similar to that described above for estimating the nonprompt-lepton background. 
The probability, \flipr, to misidentify the charge of an electron is estimated in simulation, as a function of electron \pt and $\abs{\eta}$.  
This probability ranges between $10^{-5}$ and $10^{-3}$ for electrons and is at least an order of magnitude smaller for muons.
For each SR requiring two SS leptons, a CR is constructed with the same selections except that the two leptons have opposite electric charge.  
The number of events with a lepton whose charge is misidentified is estimated by scaling each CR event by a weight obtained by summing terms of $\flipr / (1 - \flipr)$ for each lepton in the event.

Because the charge misidentification rate is estimated using simulation, the performance is evaluated in a CR enriched in $\PZ\to\Pep\Pem$ events with the charge of the electron or positron misidentified.  
For each data collection year, a single SF inclusive in \pt and $\abs{\eta}$ is derived from a comparison of the observed and estimated number of such electron pairs.  
For 2016 (2017--2018) the correction factor is approximately 1.1 (1.4).
A flat 30\% uncertainty in the estimate is assumed in all regions based on these and other studies described in the following section.

\section{Systematic uncertainties}
\label{sec:uncertainties}

Sources of systematic uncertainties related to signal and background processes and their impact on the results are summarized in Table~\ref{tab:systUnc}.
The change in SR yields is reported as a one standard deviation range.
A description of the uncertainties considered are summarized in this section, first in backgrounds estimated using CRs (charge-misidentified and nonprompt-lepton) followed by yields estimated from simulation.

The nonprompt-lepton background is assigned an uncertainty of 30\% based on the agreement observed in closure tests of the TL method using simulated QCD multijet, \ttbar, and \wjets samples. 
The simulation is used to subtract contamination from prompt leptons, which is less than 1\% in the CR, but is typically between 10\% and 20\%  in the \fr measurement control sample, with uncertainties in \fr as large as 50\% in the least statistically significant bins. 
This uncertainty leads to a 7--10\% impact on the predicted yields in the SR.
Uncertainties in the subtracted prompt contamination are considered separately for electrons and muons and correlated across bins of \fr for each lepton flavor.
The charge-misidentified lepton background is assigned an uncertainty of 30\% based on the agreement observed between the estimate and data in a control sample enriched in $\PZ\to \Pep\Pem$ events with one electron or positron having a misidentified charge. 
An uncertainty $<$5\% in the rate at which the electron charge is misidentified results from the limited number of events in the simulated samples in which this rate is measured.
The statistical uncertainty in the CRs used to estimate the charge-misidentified and nonprompt-lepton backgrounds is included in the total uncertainty.  
The statistical uncertainties in the nonprompt-lepton background and the charge-misidentified lepton background are treated as fully uncorrelated. 
The other uncertainties in these backgrounds are treated as correlated between SR bins but uncorrelated between data taking periods.
The uncertainty in the estimate of the contribution of processes with a nonprompt lepton is the dominant uncertainty in the background.

The remaining uncertainties described here apply only for the estimate of yields of the signal process and backgrounds arising from SM processes with a prompt, SS lepton pair, which are estimated using simulation. 
Statistical uncertainties originating from the finite size of simulated samples are accounted for in each SR bin and for signal and background separately~\cite{Barlow:1993dm}.
The integrated luminosities of the 2016, 2017, and 2018 data-taking periods are individually known with uncertainties in the 1.2--2.5\% range~\cite{CMS:2021xjt,CMS-PAS-LUM-17-004,CMS-PAS-LUM-18-002}, while the total 2016--2018 integrated luminosity has an uncertainty of 1.6\%, the improvement in precision reflecting the (uncorrelated) time evolution of some systematic effects.
The simulated samples are reweighted according to the observed instantaneous luminosity and the minimum bias cross section.
The uncertainty in the total inelastic \pp cross section leads to changes $<$2\% in the expected signal and background yields.

The efficiency of the trigger requirements is measured with an uncertainty of 2\%, using an independent data sample selected with single-lepton triggers. 
The efficiencies of the lepton reconstruction and identification requirements are measured using data samples enriched in $\PZ\to\Pell\Pell$ events, with uncertainties of up to 5 (3)\% per electron (muon).
Varying the lepton efficiency SFs results in a 2--3\% effect on the SR yields.
The jet energy scale is varied within its uncertainty resulting in a 1--8\% effect in signal yields and $<$5\% in background yields for rare SM processes.

The tagging efficiencies for \PQb, \PQc, and light-flavor jets are measured in dedicated data samples~\cite{Bols:2020bkb}. 
Varying these efficiencies within their measured uncertainties results in variations between 6--16\% of the SR yields. 

Uncertainties in the normalization and shape distribution of events across SRs are considered for each background process estimated from simulation.  
Flat uncertainties are applied to specific background processes to account for the uncertainty in the total cross section:5\% for $\PW\PZ$~\cite{WZxsec}, 10\% for SS $\PW\PW$~\cite{ssWWxsec}, 12\% for $\PQt\PAQt\PW$~\cite{deFlorian:2016spz}, and 8\% for $\PQt\PAQt\PZ$~\cite{deFlorian:2016spz}.
These flat uncertainties also account for the normalization changes due to the choice of the renormalization (\mur) and factorization (\muf) scales.
The uncertainties in the shape distrubtion of events for each background process from the choice of the \mur, \muf, and \alpS scales is accounted for separately and are obtained by taking the envelope of the distributions obtained with modified scales.
The typical effect of varying the \muf and \mur scales on the shapes of the BDT distributions is 10 to 15\%.
The impact of variations of the PDFs~\cite{Butterworth:2015oua} is 4 to 6\%.

For signal events, the total cross section uncertainty in the TT production mode is 6\%, estimated from the uncertainty in the \ttbar NNLO cross section, caused by the variation of the PDFs and the strong coupling \alpS~\cite{Catani:2003zt,Cacciari:2003fi,Kalogeropoulos:2018cke}.  
The uncertainty in the cross section for signal production in the ST mode is estimated to be 30\%, resulting from missing higher-order corrections.  
The total impact in the normalization of the signal yields ranges from 6 to 10\% across SRs.  
The typical effect of varying the \muf and \mur scales on the shapes of the BDT distributions is 2 to 9\% for signal processes.
The impact of variations of the PDFs~\cite{Butterworth:2015oua} is $<$2\%.

All systematic uncertainties are correlated between the SR bins.
With the exception of the \muf and \mur scales and cross section uncertainties, every systematic effect is correlated between signal and the rare SM background.
The trigger efficiency uncertainties are assumed as being uncorrelated between the data taking periods.
The uncertainty in the integrated luminosity is partially correlated between years, while all remaining uncertainties are correlated.

\begin{table*}[ht!]
\centering
\topcaption{
Sources of systematic uncertainties in the yields of signal and background processes, as well as their impact on the yields in the SRs. The impact is expressed as a one standard deviation range.}{
\cmsTable{
  \begin{scotch}{cccccc}
	\multirow{3}{*}{Source}							& Uncertainty	& Uncertainty	& Uncertainty	& Uncertainty in	& Uncertainty in \\
													& in tcH		& in tuH		& in SM SS		& nonprompt			& charge flip \\
													& prediction	& prediction	& prediction	& estimate			& estimate \\ \hline
    Estimate 										& \multirow{2}{*}{\NA} & \multirow{2}{*}{\NA} & \multirow{2}{*}{\NA} & \multirow{2}{*}{30\%} & \multirow{2}{*}{30\%} \\
	normalization									&				&				& 				& 					& \\
	\fr/\flipr								 	& \NA 			& \NA 			& \NA  			& 7--10\% 			& $<$5\% \\
	Integrated luminosity 							& 1.6\%			& 1.6\%			& 1.6\%			& \NA					& \NA \\
	Pileup 											& $<$2\% 		& $<$2\%		& $<$2\%		& \NA					& \NA \\
	Trigger efficiency 								& 2\%			& 2\%			& 2\%			& \NA					& \NA \\
	Lepton efficiency 								& 2--3\% 		& 2--3\% 		& 2--3\%		& \NA 				& \NA \\
	Jet energy scale 								& 1--6\% 		& $<$8\%		& $<$5\% 		& \NA 				& \NA \\
	\PQb{}/\PQc tagging 							& 10--16\%		& 6--13\%		& 7--14\%		& \NA 				& \NA \\
	Theory normalization 							& 6--10\% 		& 6--10\% 		& 5--25\%		& \NA 				& \NA \\
	Renormalization									& \multirow{3}{*}{7--9\%} & \multirow{3}{*}{2--6\%} & \multirow{3}{*}{10--15\%} & \multirow{3}{*}{\NA} & \multirow{3}{*}{\NA} \\
	and	factorization  								&				&				&				&					& \\
	scale shape 									& 				&				&				&					& \\
	PDF shape 										& $<$2\%		& $<$2\%		& 4--6\%		& \NA 				& \NA \\
	\hline
	Total 											& 14--16\% 		& 11--14\%		& 20--28\%		& 31--35\%			& 29--31\% \\
  \end{scotch}
}
}
\label{tab:systUnc}
\end{table*}

\section{Results}
\label{sec:results}

\begin{figure*}[ht!]
\centering
        \includegraphics[width=0.45\textwidth]{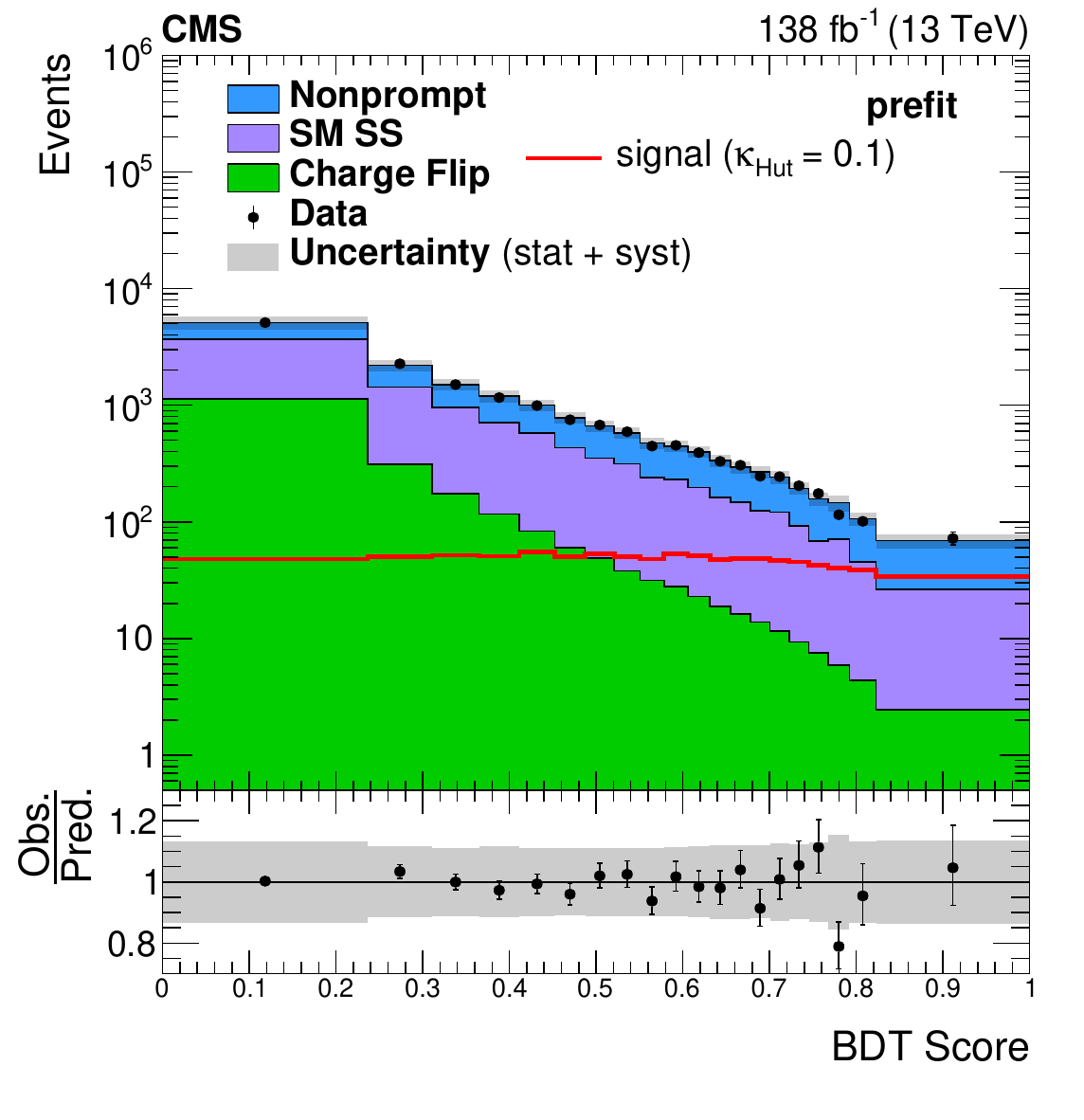}
        \includegraphics[width=0.45\textwidth]{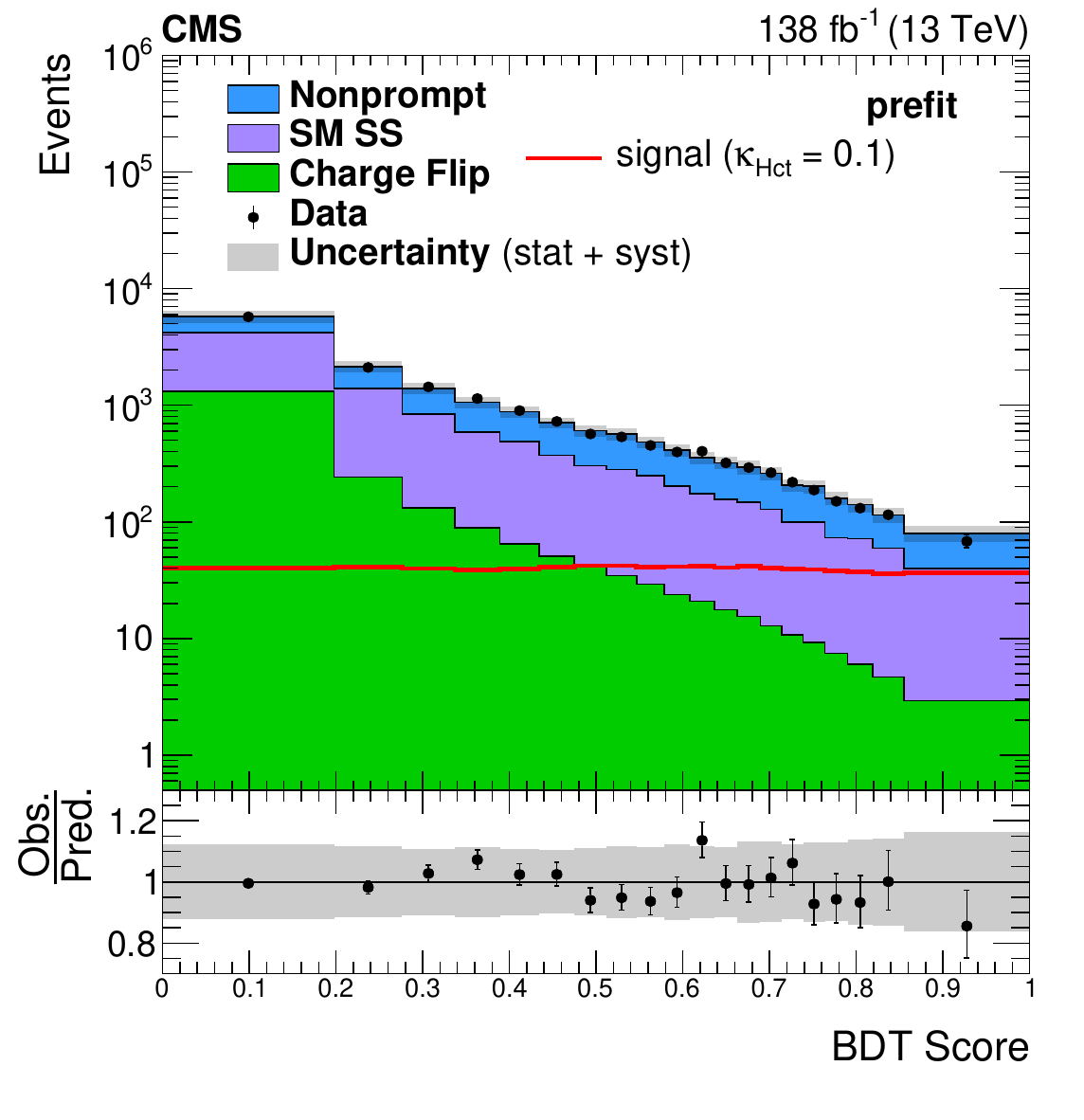}\\
        \includegraphics[width=0.45\textwidth]{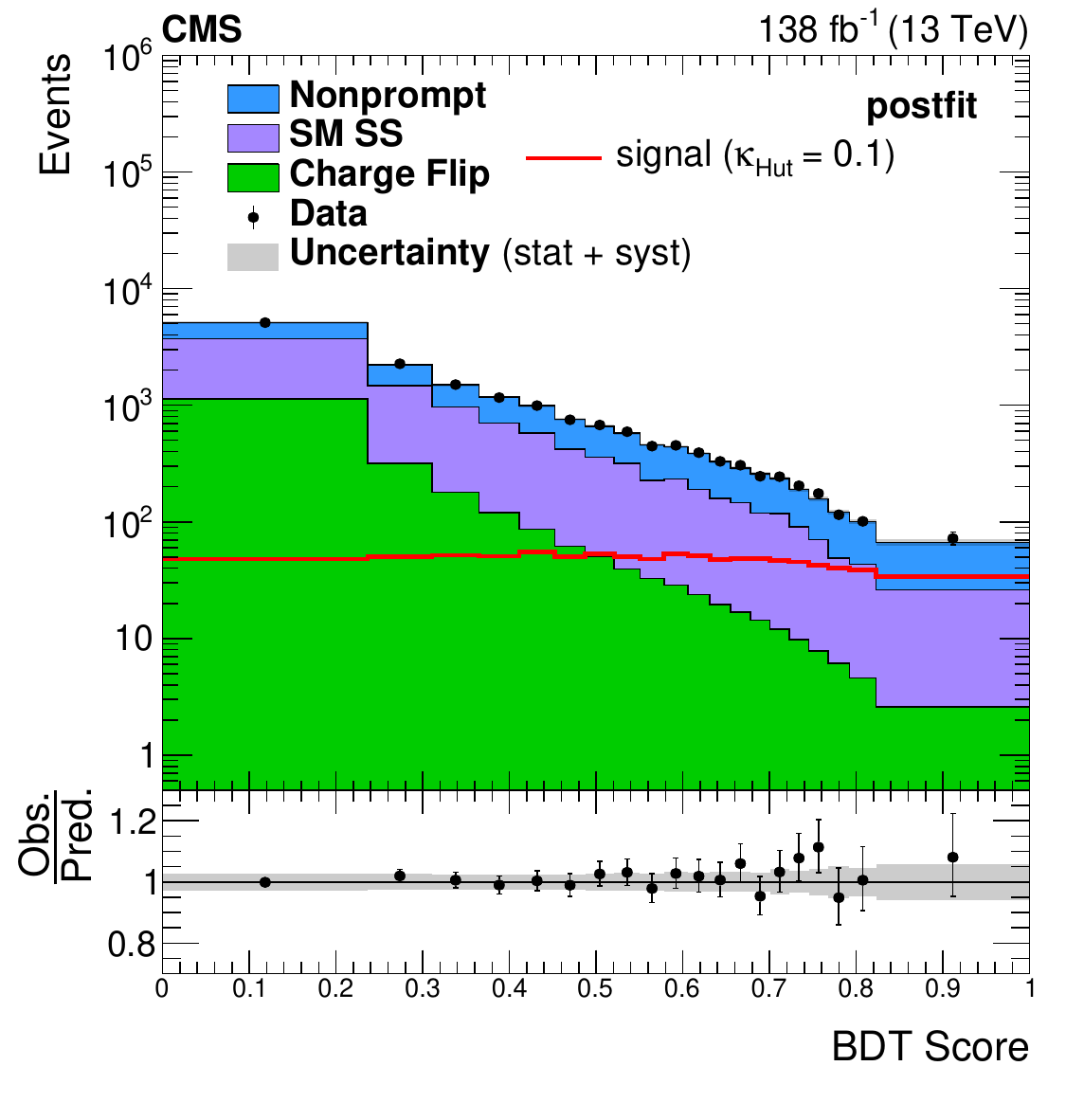}
        \includegraphics[width=0.45\textwidth]{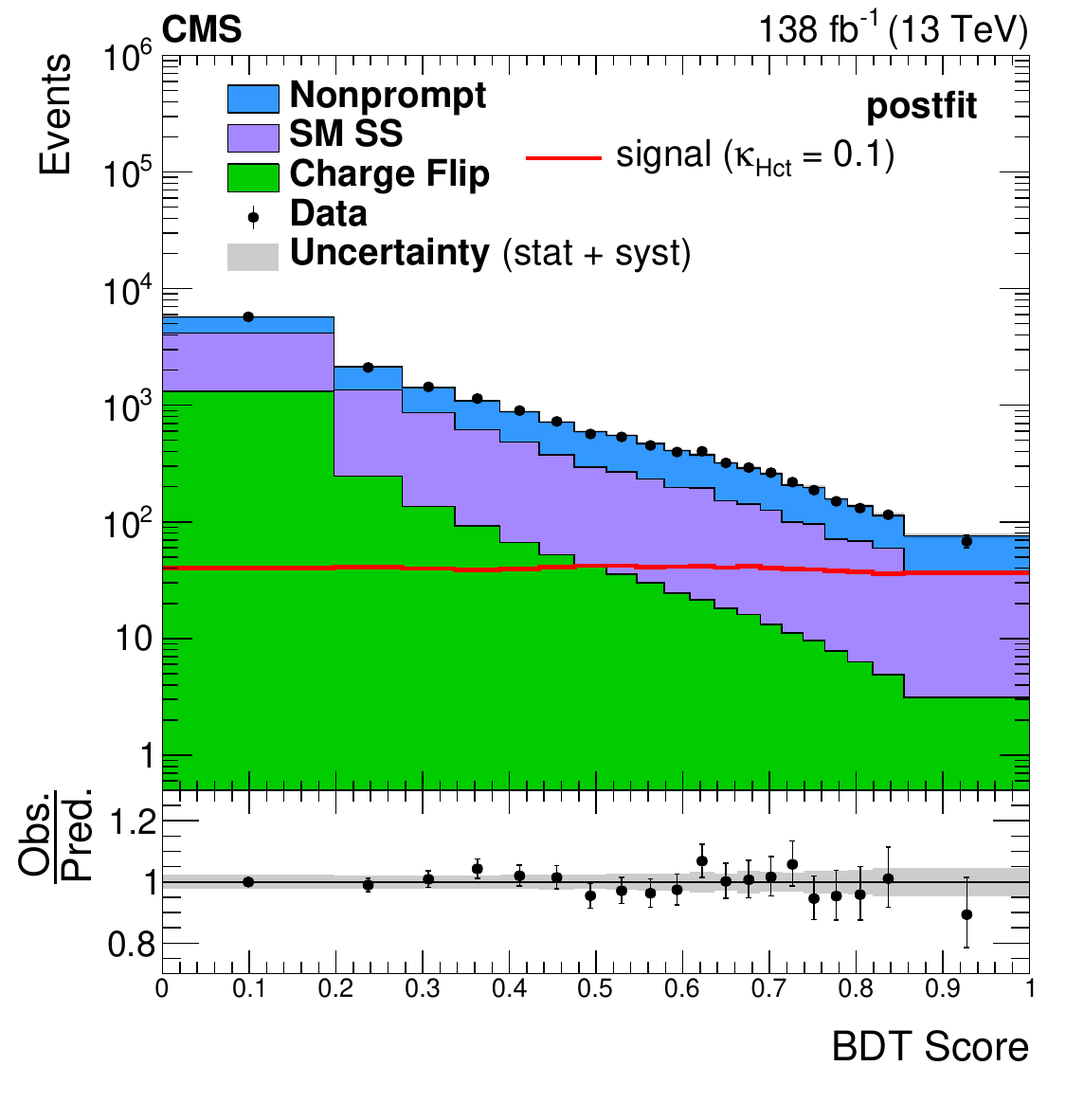}
        \caption{The prefit (upper) and postfit (lower) observed and expected distributions of BDT score in the SR for 2016-2018 data are shown. The \tuH signal is shown on the left and \tcH is shown on the right, normalized with a coupling $\kappa_{\hqt}=0.1$ for illustration. The uncertainty bands include statistical and systematic uncertainties in the estimated backgrounds.}
        \label{fig:postfitDist}
\end{figure*}

A binned likelihood fit is constructed using yields from the signal and control regions, incorporating the theoretical and experimental uncertainties described in Section~\ref{sec:uncertainties} as nuisance parameters.  
The number of observed events and of simulated events in the SRs and CRs are Poisson distributed. 
Expected yields of signal events are estimated using a mixture of simulated samples of TT and ST production modes weighed according to the cross sections in Section~\ref{sec:mc}.
Figure~\ref{fig:postfitDist} shows that the estimated backgrounds and observed data are consistent within statistical and systematic uncertainties as a function of the BDT score, before (``prefit'') and after maximizing the likelihood function (``postfit''). 
Knowledge of the backgrounds, particularly the non-prompt background, is improved considering the signal and control region yields in the fit.
Overlaid are distributions of signal, normalized assuming a coupling strength of 0.1 between the top quark, the Higgs boson, and an up or charm quark.
While the presence of FCNC interactions would affect other Higgs boson couplings, existing experimental constraints on FCNC interactions limit the potential impact on the Higgs boson decay modes, rendering the effects negligible.

\begin{figure}[ht!]
\centering
	\includegraphics[width=0.4\textwidth]{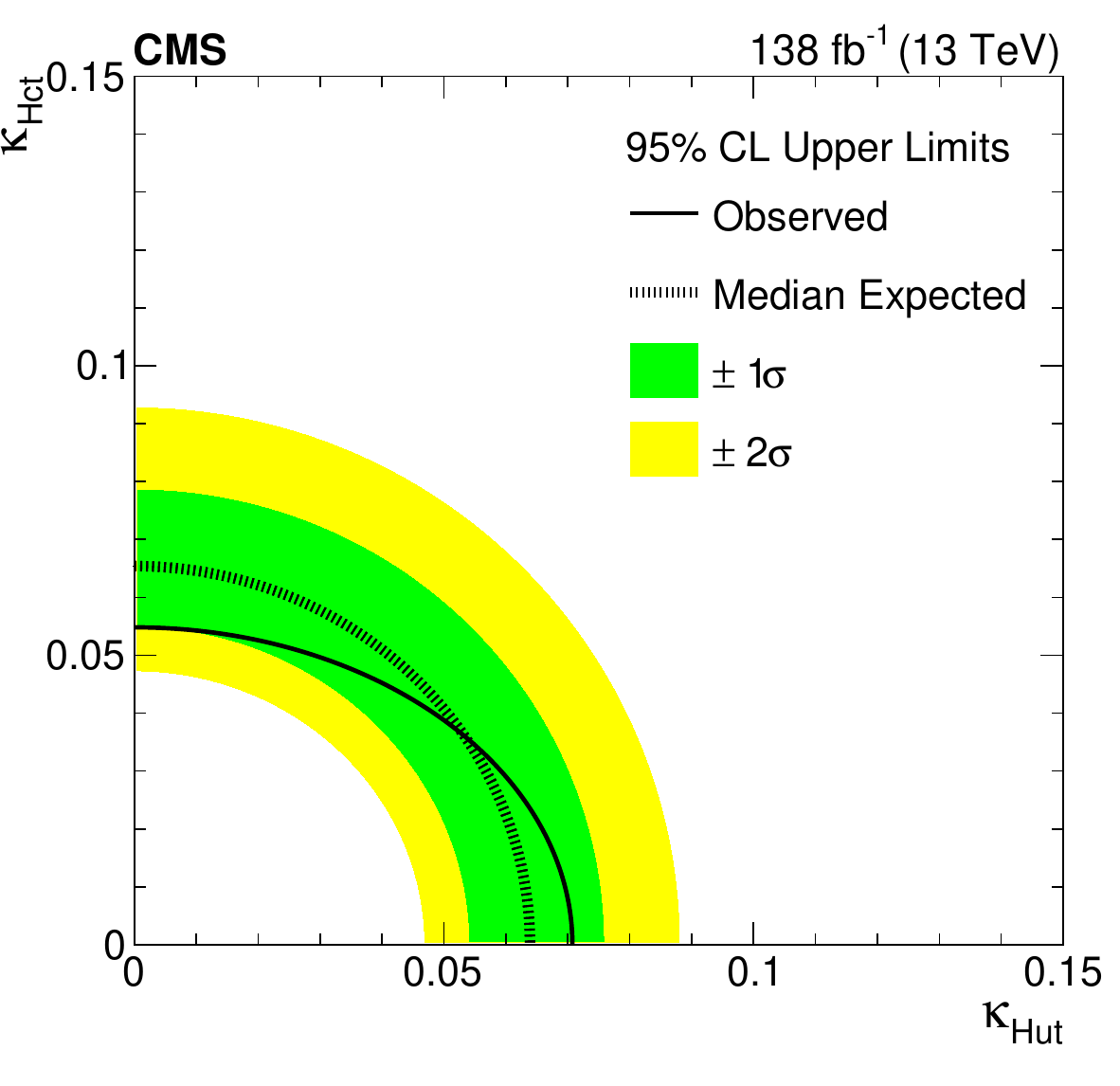}
	\includegraphics[width=0.4\textwidth]{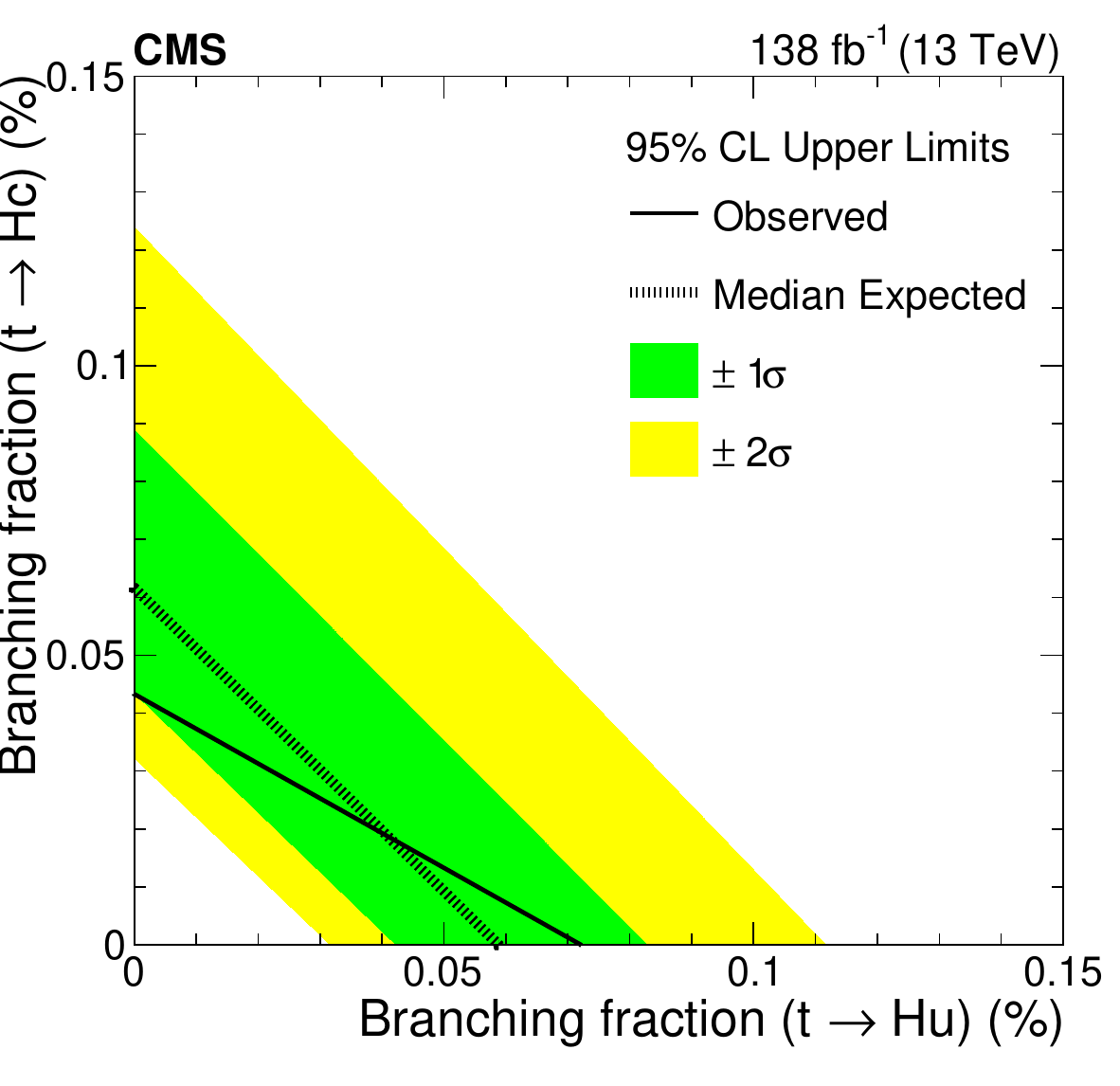}
	\caption{The expected (dashed line) and observed (solid line) limits on the coupling strength (\cmsLeft) and on the branching fraction (\cmsRight) are shown.  The green (yellow) bands show the 68 (95)\% confidence level ranges of the expected limits. The area to the right and above the solid line is excluded.}
	\label{fig:results}
\end{figure}

Results are obtained using the CMS statistical analysis tool~\textsc{Combine}~\cite{CAT-23-001} which is based on the RooFit~\cite{Verkerke:2003ir} and RooStats~\cite{Moneta:2010pm} frameworks.
Upper limits on the coupling strengths and branching fractions are set
at 95\% confidence level (\CL), using the \CLs criterion with the LHC profile likelihood ratio as a test statistic in the asymptotic formulation~\cite{Junk_1999, Read_2002,Cowan:2010js,ATLAS:2011tau}.
Observed (expected) limits are 0.072 (0.059)\% on the \brHut and 0.043 (0.062)\% on the \brHct.
The limits can be cast as observed (expected) constraints on the anomalous coupling strengths:  $\kappa_{\hut} < 0.071 (0.064)$ and $\kappa_{\hct} < 0.055 (0.065)$.
The limits agree within one standard deviation, see Fig.~\ref{fig:results}, 
which shows the expected and observed limits, as well as the one and two standard deviation bands on the coupling strength and on the branching fraction for each anomalous coupling.
The exclusion curves shown in Fig.~\ref{fig:results} are obtained by interpolating the limits with only a single coupling present, assuming a linear relationship in the branching fractions.

\section{Combination}
\label{sec:combination}

\begin{figure}[ht!]
	\centering
	\includegraphics[width=.49\textwidth]{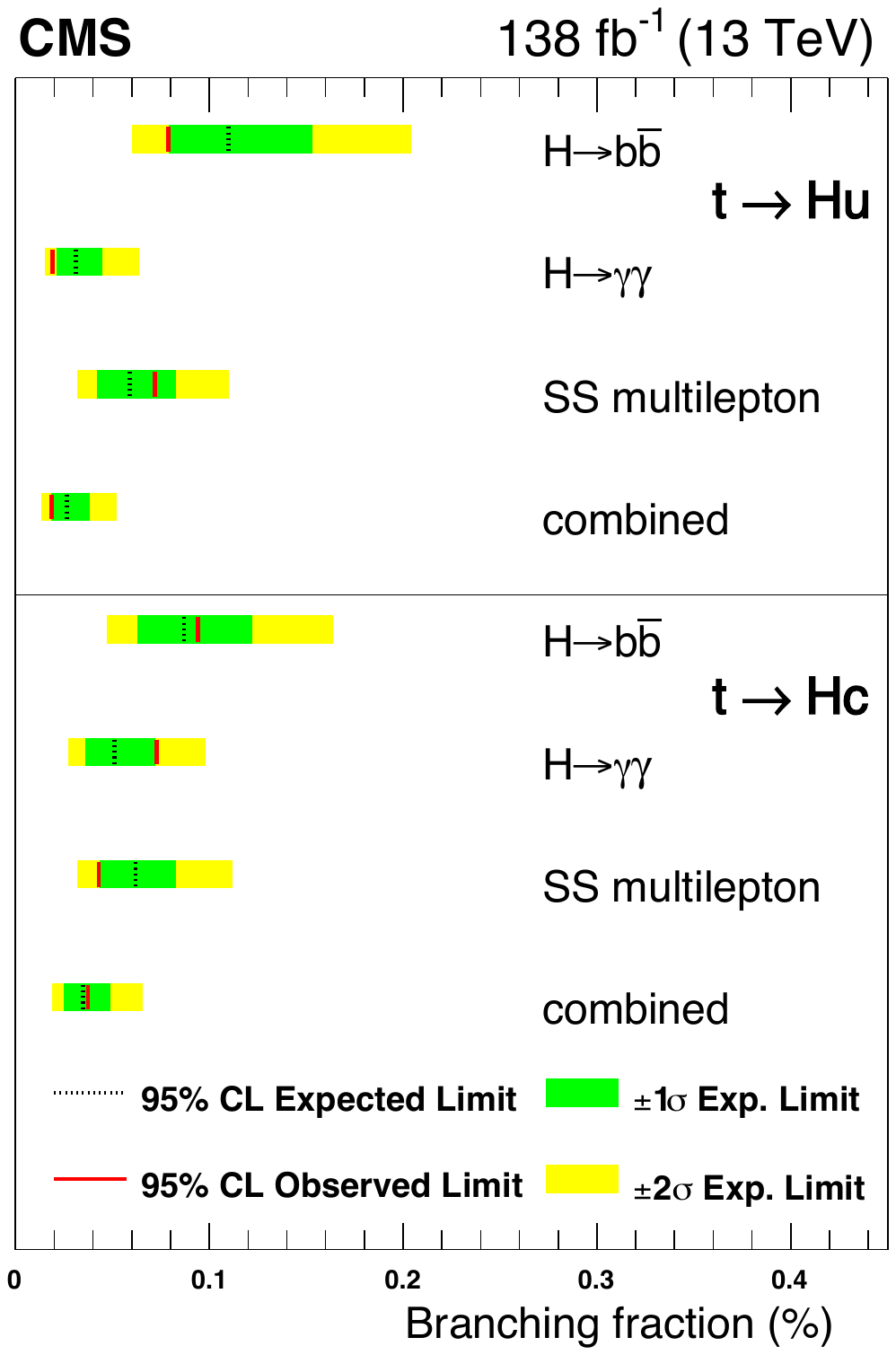}
	\caption{Summary of the observed and expected results from the three individual analyses and their combination. The dotted line shows the expected limits and the solid red lines shows the observed limits. The green and yellow bands show the 68 and 95\% confidence level ranges of the expected limits.}
	\label{fig:comb_summary}
\end{figure}

In addition to the results presented in the previous section, the CMS Collaboration has reported the results of two other searches for Higgs-mediated FCNC decays of the top quark, in which the Higgs boson decays to bottom quark pairs~\cite{CMS:2021gfa} or to a pair of photons~\cite{CMS:2021cqc}.  
The results of these searches and the new search presented are summarized in Table~\ref{tab:comb_ind_results}.  
A statistical combination of the results of the three searches is performed.  
As the three results target different and distinct decay modes of the Higgs boson, there is no overlap between the SRs of the three analyses.  
Correlations of systematic uncertainties in the signal and background yields are studied and accounted for.
Uncertainties are correlated between the three analyses if the object definitions and approaches to measuring uncertainties are the same; otherwise, they are uncorrelated.
Uncertainties in the jet energy scale and \ptmiss resolution, 
the integrated luminosity, the lepton identification efficiencies, and per-process theoretical uncertainties are treated as fully correlated.  
Remaining uncertainties are treated as uncorrelated.  
The impact of the choice of the correlation scheme on the combination result was studied and found to be $<$1\%.
The results of the combination are reported in Table~\ref{tab:comb_ind_results}. 
A summary and comparison of the results of the individual analyses and the combination is shown in Fig.~\ref{fig:comb_summary}.

\begin{table*}[ht!]
	\centering
	\topcaption{Observed and expected limits on the \tuH and \tcH branching fractions for the three searches in different Higgs boson decay channels performed by the CMS Collaboration. A statistical combination of the results is also reported.}
	\label{tab:comb_ind_results}
	\begin{scotch}{ccc}
        \multirow{2}{*}{Analysis} & 95\% CL limit on \brHut & 95\% CL limit on \brHct \\
                                  & observed (expected) & observed (expected) \\
	\hline
	$\PH\to\PQb\PAQb$~\cite{CMS:2021gfa}	 & 0.079 (0.11)\%	& 0.094 (0.086)\%	\\
	$\PH\to\PGg\PGg$~\cite{CMS:2021cqc}			 & 0.019 (0.031)\%	& 0.073 (0.051)\%	\\
	Leptonic  (this analysis)		 		& 0.072 (0.059)\%	& 0.043 (0.062)\%   \\                                                     
    	Combination          	 				& 0.019 (0.027)\%  	& 0.037 (0.035)\% 
	\end{scotch}
\end{table*}

\section{Summary}
\label{sec:summary}

This paper presents the results of a search for flavor changing neutral current interactions of the top quark (\PQt), Higgs boson (\PH), and an up (\PQu) or charm (\PQc) quark. The search is performed in a final state with a pair of leptons of same electric charge and at least one jet. 
Expected yields from backgrounds emerging from detector effects are estimated by extrapolating yields observed in control regions using transfer factors measured in orthogonal data or simulated samples.  
Expected yields from standard model processes producing a pair of prompt leptons with the same-sign electric charge are estimated from simulation.  
Two trained boosted decision trees are used to evaluate and classify each event.
No excess above the estimated background from standard model processes is observed. 
The observed (expected) upper limits at 95\% confidence level (\CL) on the branching fraction are found to be 0.072\% (0.059\%) for $\mathcal{B}(\tuH)$ and 0.043\% (0.062\%) for $\mathcal{B}(\tcH)$.
The limits can be cast as observed (expected) constraints on the anomalous coupling strengths:  $\kappa_{\hut} < 0.071 (0.064)$ and $\kappa_{\hct} < 0.055 (0.065)$. 

The ATLAS Collaboration has performed a similar search, including a targeted $\PH\rightarrow\PGtp\PGtm$ final state~\cite{ATLAS:2024mih}. A combination of ATLAS searches in different final states obtained observed (expected) limits at 95\% CL of $\mathcal{B}(\tuH) < 0.026\%(0.018\%)$ and  $\mathcal{B}(\tcH) < 0.034\%(0.023\%)$.
A statistical combination of the results of the search in leptonic final states with those of previous CMS publications searching for the same phenomena where the Higgs boson decays to a bottom quark-antiquark pair or to a pair of photons is performed.
The results of the combination lead to observed (expected) exclusion limits at the 95\% \CL on the branching fractions $\mathcal{B}(\tuH)< 0.019\% (0.027\%)$ and  $\mathcal{B}(\tcH)< 0.037\% (0.035\%)$ and represent the most stringent constraints on these interactions from the CMS Collaboration. 

\begin{acknowledgments}
	We congratulate our colleagues in the CERN accelerator departments for the excellent performance of the LHC and thank the technical and administrative staffs at CERN and at other CMS institutes for their contributions to the success of the CMS effort. In addition, we gratefully acknowledge the computing centers and personnel of the Worldwide LHC Computing Grid and other centers for delivering so effectively the computing infrastructure essential to our analyses. Finally, we acknowledge the enduring support for the construction and operation of the LHC, the CMS detector, and the supporting computing infrastructure provided by the following funding agencies: SC (Armenia), BMBWF and FWF (Austria); FNRS and FWO (Belgium); CNPq, CAPES, FAPERJ, FAPERGS, and FAPESP (Brazil); MES and BNSF (Bulgaria); CERN; CAS, MoST, and NSFC (China); MINCIENCIAS (Colombia); MSES and CSF (Croatia); RIF (Cyprus); SENESCYT (Ecuador); ERC PRG, RVTT3 and MoER TK202 (Estonia); Academy of Finland, MEC, and HIP (Finland); CEA and CNRS/IN2P3 (France); SRNSF (Georgia); BMBF, DFG, and HGF (Germany); GSRI (Greece); NKFIH (Hungary); DAE and DST (India); IPM (Iran); SFI (Ireland); INFN (Italy); MSIP and NRF (Republic of Korea); MES (Latvia); LMTLT (Lithuania); MOE and UM (Malaysia); BUAP, CINVESTAV, CONACYT, LNS, SEP, and UASLP-FAI (Mexico); MOS (Montenegro); MBIE (New Zealand); PAEC (Pakistan); MES and NSC (Poland); FCT (Portugal);  MESTD (Serbia); MCIN/AEI and PCTI (Spain); MOSTR (Sri Lanka); Swiss Funding Agencies (Switzerland); MST (Taipei); MHESI and NSTDA (Thailand); TUBITAK and TENMAK (Turkey); NASU (Ukraine); STFC (United Kingdom); DOE and NSF (USA).

	\hyphenation{Rachada-pisek} Individuals have received support from the Marie-Curie program and the European Research Council and Horizon 2020 Grant, contract Nos.\ 675440, 724704, 752730, 758316, 765710, 824093, 101115353, 101002207, and COST Action CA16108 (European Union); the Leventis Foundation; the Alfred P.\ Sloan Foundation; the Alexander von Humboldt Foundation; the Science Committee, project no. 22rl-037 (Armenia); the Belgian Federal Science Policy Office; the Fonds pour la Formation \`a la Recherche dans l'Industrie et dans l'Agriculture (FRIA-Belgium); the F.R.S.-FNRS and FWO (Belgium) under the ``Excellence of Science -- EOS" -- be.h project n.\ 30820817; the Beijing Municipal Science \& Technology Commission, No. Z191100007219010 and Fundamental Research Funds for the Central Universities (China); the Ministry of Education, Youth and Sports (MEYS) of the Czech Republic; the Shota Rustaveli National Science Foundation, grant FR-22-985 (Georgia); the Deutsche Forschungsgemeinschaft (DFG), among others, under Germany's Excellence Strategy -- EXC 2121 ``Quantum Universe" -- 390833306, and under project number 400140256 - GRK2497; the Hellenic Foundation for Research and Innovation (HFRI), Project Number 2288 (Greece); the Hungarian Academy of Sciences, the New National Excellence Program - \'UNKP, the NKFIH research grants K 131991, K 133046, K 138136, K 143460, K 143477, K 146913, K 146914, K 147048, 2020-2.2.1-ED-2021-00181, and TKP2021-NKTA-64 (Hungary); the Council of Science and Industrial Research, India; ICSC -- National Research Center for High Performance Computing, Big Data and Quantum Computing and FAIR -- Future Artificial Intelligence Research, funded by the NextGenerationEU program (Italy); the Latvian Council of Science; the Ministry of Education and Science, project no. 2022/WK/14, and the National Science Center, contracts Opus 2021/41/B/ST2/01369 and 2021/43/B/ST2/01552 (Poland); the Funda\c{c}\~ao para a Ci\^encia e a Tecnologia, grant CEECIND/01334/2018 (Portugal); the National Priorities Research Program by Qatar National Research Fund; MCIN/AEI/10.13039/501100011033, ERDF ``a way of making Europe", and the Programa Estatal de Fomento de la Investigaci{\'o}n Cient{\'i}fica y T{\'e}cnica de Excelencia Mar\'{\i}a de Maeztu, grant MDM-2017-0765 and Programa Severo Ochoa del Principado de Asturias (Spain); the Chulalongkorn Academic into Its 2nd Century Project Advancement Project, and the National Science, Research and Innovation Fund via the Program Management Unit for Human Resources \& Institutional Development, Research and Innovation, grant B39G670016 (Thailand); the Kavli Foundation; the Nvidia Corporation; the SuperMicro Corporation; the Welch Foundation, contract C-1845; and the Weston Havens Foundation (USA).	
\end{acknowledgments}

\bibliography{auto_generated} 
\cleardoublepage \appendix\section{The CMS Collaboration \label{app:collab}}\begin{sloppypar}\hyphenpenalty=5000\widowpenalty=500\clubpenalty=5000
\cmsinstitute{Yerevan Physics Institute, Yerevan, Armenia}
{\tolerance=6000
A.~Hayrapetyan, A.~Tumasyan\cmsAuthorMark{1}\cmsorcid{0009-0000-0684-6742}
\par}
\cmsinstitute{Institut f\"{u}r Hochenergiephysik, Vienna, Austria}
{\tolerance=6000
W.~Adam\cmsorcid{0000-0001-9099-4341}, J.W.~Andrejkovic, T.~Bergauer\cmsorcid{0000-0002-5786-0293}, S.~Chatterjee\cmsorcid{0000-0003-2660-0349}, K.~Damanakis\cmsorcid{0000-0001-5389-2872}, M.~Dragicevic\cmsorcid{0000-0003-1967-6783}, P.S.~Hussain\cmsorcid{0000-0002-4825-5278}, M.~Jeitler\cmsAuthorMark{2}\cmsorcid{0000-0002-5141-9560}, N.~Krammer\cmsorcid{0000-0002-0548-0985}, A.~Li\cmsorcid{0000-0002-4547-116X}, D.~Liko\cmsorcid{0000-0002-3380-473X}, I.~Mikulec\cmsorcid{0000-0003-0385-2746}, J.~Schieck\cmsAuthorMark{2}\cmsorcid{0000-0002-1058-8093}, R.~Sch\"{o}fbeck\cmsorcid{0000-0002-2332-8784}, D.~Schwarz\cmsorcid{0000-0002-3821-7331}, M.~Sonawane\cmsorcid{0000-0003-0510-7010}, S.~Templ\cmsorcid{0000-0003-3137-5692}, W.~Waltenberger\cmsorcid{0000-0002-6215-7228}, C.-E.~Wulz\cmsAuthorMark{2}\cmsorcid{0000-0001-9226-5812}
% ...
\par}
\vskip\cmsinstskip
\dag:~Deceased\\
$^{1}$Also at Yerevan State University, Yerevan, Armenia\\
$^{2}$Also at TU Wien, Vienna, Austria\\
% ...
\end{sloppypar}
\end{document}